# MG1-688432: A Peculiar Variable System[1]

ROY A. TUCKER[*]
*Goodricke-Pigott Observatory, 5500 West Nebraska Street, Tucson, AZ 85757; GNAT, Inc., Tucson, AZ 85745;*

ERIC R. CRAINE[2]
*Western Research Company, Inc., 3275 W. Ina Road, Ste. 221-A1, Tucson, AZ 85741; GNAT, Inc., Tucson, AZ 85745;*

BRIAN L. CRAINE
*Western Research Company, Inc., 85 Bolinas Road, Ste. 18, Fairfax, CA 94930; GNAT, Inc., Tucson, AZ 85745*

ANDY S. KULESSA
*Colin Gum Observatory, Greenhill, SA, Australia; GNAT, Inc., Tucson, AZ 85745*

CHRISTOPHER J. CORBALLY
*Vatican Observatory Research Group, Steward Observatory, University of Arizona, Tucson, AZ 85721*

ADAM L. KRAUS
*Department of Astronomy, University of Texas, Austin, TX 78712*

[*] We regret to report the passing (2021 Mar 5) of our friend and colleague Roy A. Tucker.

**Abstract**

The short period variable star MG1-688432 has been discovered to exhibit occasional extremely high energy optical outbursts as high as $10^{38}$ ergs. Outbursts are typically of several hours duration. These events are often highly structured, resembling sequential associated releases of energy. Twenty years of time sequence photometry is presented, indicating a basic sinusoidal light curve of mean period 6.65d, with some phase shifting and long-term temporal trends in amplitude and mean brightness. Spectroscopy reveals a peculiar star, best resembling a K3 subgiant that has evolved off the main sequence moderately red-ward of the giant branch. Spectroscopic and radial velocity analyses indicate a binary system orbiting its barycenter with an unseen companion to the K3IV primary. This is not an eclipsing system, with the inclination of the orbit precluding eclipse by the secondary. The system is at a distance of 1.5kpc and analysis of Gaia observations leads to the conclusion that the HR diagram position of MG1-688432 is established by an intrinsic feature of the system, most likely either the stellar evolutionary state of the observed star or the presence of small (non-gray) dust within the system. Two mechanisms (or combinations thereof) that might give rise to characteristics of the system are 1) magnetically induced chromospheric activity, and 2) impacts with tidally disrupted planetary debris.

# 1. Introduction

The Moving Object and Transient Event Search System (MOTESS) originated as a very successful asteroid discovery project of the Goodricke-Pigott Observatory (GPO) in Tucson, Arizona, in the form of a long term, systematic temporal imaging survey of the celestial equator region (cf. Tucker 2007). Subsequently, there was a collaborative effort between GPO and the Global Network of Astronomical Telescopes, Inc. of Tucson, Arizona (GNAT) in which GPO provided raw MOTESS sky survey image data and GNAT developed a data reduction pipeline for photometric measurement of all stars detected in the images, and the subsequent extraction of interesting variable star candidates.

The star MG1-688432 was characterized as a photometric variable star with the publication of the first MOTESS-GNAT Variable Star Catalog (MG1-VSC) (Kraus et al. 2007). Because it is relatively bright (between $R_{MG}$ about 13.7 – 14.2 mag), it has been listed, under numerous names, in a number of astrometric, proper motion, and various survey catalogs, the most relevant of which will be referenced throughout.

Examination of the statistics and discovery light curves of stars in the MG1-VSC has proven a rich source of interesting candidates for various types of follow-up analyses (cf. Tucker et al. 2013; Craine et al. 2013) and additional observation (cf. Kraus et al. 2011). It was such an effort that initially drew attention to MG1-688432, which exhibited sinusoidal variability in the MOTESS light curves with an amplitude of 0.5 magnitudes and an apparent period (albeit with systematic uncertainty due to aliasing) of P ~ 6.7 days. Follow-up photometric observations made to confirm and refine the period of the star serendipitously, and quickly, led to discovery of a highly energetic, short duration outburst that suggested this was an interesting star, which in turn led to the additional observations and analysis reported here.

This paper announces the discovery of multiple modes of photometrically variable behavior of MG1-688432, summarizes the observations that have been made to date, outlines the data reduction protocols, and discusses the results of those observations. Also presented are what are thought to be the most plausible models to describe this star, and continuing observations that may be both interesting and illuminating in the validation (or refutation) of these models.

# 2. Observations
## 2.1. Instrumentation

The source images were obtained using the MOTESS time delay integration (TDI) telescopes of GPO in Tucson, Arizona as described in detail by Tucker (2007). MOTESS produces three images, separated in time by about 20-min, of each field in the survey strip on every clear night of observation. This procedure allowed for discovery of asteroids that drifted through the fields of each three-image set. This was to later become a valuable tool for confirmation of temporal brightness behavior in field stars as well.

Basic parameters of MOTESS are as follows:

- CCD – 1024 × 1024 SITe TK1024 device with 24-micron pixels.
- Integration time – 191.488 seconds (for MG1 only, set by CCD dimensions, optics, field declination, and earth's rotation).
- Field of view – 0.8 degree wide strip extending 120 to 200 degrees, depending upon observing season.
- Image scale – 2.83 arcseconds per pixel.

Follow-up observations of the MOTESS-GNAT variable star candidates are typically conducted with a wide range of distributed small telescopes. In the case of MG1-688432, these observations were made using an 8-in Schmidt Cassegrain Telescope (SCT) and 14-in SCT (both by Celestron) and SBIG cameras (ST-7, ST-8, or ST-9) at GPO. The star field was usually imaged with an open channel (clear filter), but also through Bessell B, V, R, and I-band filters at various times during the follow-up period beginning in 2010. The camera data acquisition was controlled by MaximDL software (Diffraction Limited: Ottawa, Ont. Canada). The mode of observation was to acquire the target field, autoguide on a suitable guide star, set the integration time, and invoke continuous sequential imaging until the end of the usable night. Darks were also obtained each night of observing with periodic twilight flats that were found to be very stable due to the static configuration of telescopes and cameras.

Follow-up photometry was also conducted at Colin Gum Observatory (CGO) in Greenhill, South Australia using a 12-in SCT by Meade, equipped with an SBIG STT-8300M camera. The detector is a Kodak KAF-8300 CCD, featuring a pixel array of 3326×2504px. The maximum quantum efficiency of 56% occurs at a wavelength of 550nm. Observations were made with both clear channel and a Johnson V filter.

Spectroscopy was obtained on three occasions in 2012 and 2013 at the 2.23-m Bok telescope (University of Arizona, Kitt Peak Station) using the Boller & Chivens spectrograph with the 400 line/mm, 6.30 degree blaze grating, yielding a dispersion of 2.7578 Angstroms/pixel, and the 1200 line/mm, 21.10 degree blaze grating, yielding a dispersion of 0.97 Angstroms/pixel onto a 1200×800 CCD with 15 micron square pixels. The camera was typically set to provide vertical binning of two or four during the observations. When using the 400 line/mm grating, the spectrograph was typically set to provide coverage from about 3600 Å to about 6900 Å. When using the 1200 line/mm grating, the instrument was set to provide coverage from about 6300 Å to about 7450 Å.

A week of spectroscopic observation was made in 2018 Apr at the 1.8-m Vatican Advanced Technology Telescope (VATT) of the Vatican Observatory (VO) using the VATT Spectrograph with a 600 line/mm grating, variously tilted to optimize either the blue or the red end of the visual spectrum. Additional VATT spectroscopy was obtained over six nights with the same spectrograph (600 line/mm) in the red region of the spectrum in 2020 Feb.

## 2.2. Inventory of Observations
### 2.2.1. MG1 and MG6 Photometry

The MG1-VSC (Kraus et al. 2007) yielded 247 open channel photometric observations of MG1-688432 obtained during the 2001 Apr to 2003 Jul period of the first MG Survey. V and I bandpass images were occasionally obtained during bright moon. These observations were obtained with two of the three telescopes in the MOTESS suite of instruments. The observations took the form of a collection of FITS format images of 48.3 arcmin$^2$ field of view. The integration time for the 35cm aperture telescope/CCD camera system fields, operated in time delay integration mode, was 191.488 sec.

The MG6 sky survey, conducted similarly to the MG1 survey, for the same declination as MG1, but during the period 2011 Aug to 2013 May, has now been reduced, but not yet published. MG6 yielded light curve data for MG1-688432 similar to that of MG1 but ten years after the original discovery observations.

*2.2.2. Follow-up Photometry*

Follow-up pointed photometry made at GPO consists of 41,805 observations made during 699 nights over the period 2010 - 2020. This was augmented by 348 observations made during 37 nights in 2018 – 2020 at CGO. Many of the latter observations were simultaneous with nights of spectroscopic observing at the VATT.

*2.2.3. Spectroscopy*

A total of 218 spectra were obtained of MG1-688432 over 15 nights of observing on two telescopes. Additional spectra were made at the VATT of a flux calibration star 14 CVn. The follow-up photometry comparison star was also observed spectroscopically during the 2020 Feb VATT observing run, and was found to be a K3-4V star, very similar in color to MG1-688432 which was found to be a K3 subgiant.

### 3. Data Reduction

*3.1. MG1-VSC*

Data reduction techniques for the MG1-VSC have been described in detail elsewhere (Kraus et al. 2007). Some follow-up reduction of MG1-VSC data for specific individual stars takes the form of attempted refinement of the periodogram analysis of MG1 light curves. The period analysis of the data in the MG1-VSC is only a rough indicator of periodicity and periods of the cataloged stars, and often has large errors. These errors arise for several reasons, including the cadence-related aliasing of short periods, and the uncertainty of variable type. For example, true eclipsing binaries do not yield their correct periods to the MG1-VSC periodogram algorithm. In spite of these known errors, it was decided to include the calculated periods and resultant phased light curves for the simple reason that the phased curves give a clear visual impression of which variable stars are likely to be periodic at some level. An effort was made in the MG1-VSC to warn the user about the tentative nature of many of the periods by including a "false alarm" metric in the cataloged data. This metric is often ignored, but one does so at one's own peril! In the case of the original MG1 data, the photometric observations of MG1-688432 were re-analyzed with the Peranso software utility (www.peranso.com).

*3.2. MG1 and MG6*

The MG1 and MG6 surveys, both containing MG1-688432, were recently processed using a newly developed internal data pipeline and resulting in a relational database rather than a fixed variable star catalog. The new data reduction protocol is described briefly here.

*3.2.1. Image pre-processing*

The raw FITS formatted images were pre-processed by night and telescope using a Python MOTESS-GNAT Image Reduction Script (pyMG-IRS) operating in a 32-bit environment on a Windows desktop computer. The dark frames were automatically identified by the image median intensity. Ten of the science images with below mean average intensities were selected for creation of synthetic flat fields (Tucker 2007). The science images were corrected for dark field and flat fielding and then each image was overlapped with 256 lines of the successive image to insure no loss of data between fields. The image was then plate-solved by calling a Pinpoint DLL (DC-3 Dreams, Mesa, AZ) with reference to the UCAC 4.0 database (Zachrias et al. 2013). It was the use of the Pinpoint functions that restricted the process to a 32-bit environment. Images that did not plate solve, or that were too bright to be useful, were rejected at this stage.

### 3.2.2. Photometry

Photometry was realized using the Python MOTESS-GNAT Survey Photometry and Cataloging Script (pyMG-SPCS) operating in a 64-bit environment and making heavy use of the Astropy/Photutils library (Astropy Collaboration 2018; Bradley et al. 2019). The 64-bit environment was key to this analysis by allowing the maintenance of the large arrays of generated data in memory during the processing of the entire survey.

The pre-processed images were qualified for acquisition between astronomical twilight and sun rise, background lighting evenness, and high background level. The remaining images were background subtracted using a custom local median (32×32 pixel) subtract routine. Objects were detected using the Astropy DAOStarFind routine.

Fixed aperture photometry was then performed using the Photutils *aperture_photometry* function. The returned objects were limited by a minimum and maximum allowed aperture sum to improve measurement precision and linearity. Due to the overlap region of images any one telescope per night will have duplicated measurements for some objects. The duplicate objects were deleted, and the remaining detected objects added to a master array of objects either as new objects or, if already detected, as a new measurement of an object.

The differential magnitude was determined using an ensemble approach. Ensemble intensities were determined as the average intensity for the objects in successive 3-min RA blocks and the differential magnitude determined using the block corresponding to the position of the object. Nightly paired magnitudes (i.e., from telescope A and telescope B), acquired about 20 minutes apart, were required to match within 0.2 magnitude or they were discarded. The object's right ascension and declination, observed differential magnitudes, and derived metrics were then written to an SQL database.

### 3.3. Follow-up Photometry

Follow-up photometry was reduced using standard techniques with the MaximDL software utility. Raw FITS images were dark subtracted and flat fielded, and the program star, a comparison star, and a check star were measured using a software aperture and the MaximDL algorithms for determining instrumental magnitudes. The same comparison and check stars were used throughout the program for all of the follow-up observations made by the authors of this paper. The ensemble mean comparison star protocols for MG1 and MG6 are described in Section 3.2.2. The follow-up differential photometry was scaled relative to the known magnitude of the constant check star to ensure that consistent zero-point offsets were applied throughout the entire follow-up dataset.

In the case of photometry performed at Colin Gum Observatory, short integrations dictated by limitations in the guiding system were co-added to produce images in which the signal-to-noise ratio (SNR) approximated that of the Goodricke-Pigott Observatory observations.

### 3.4. Spectroscopy

VATT spectroscopy was reduced using standard protocols in the MaximDL software utility. Dark subtraction was followed by median combining several images to remove radiation events and high frequency shot noise. Flat-fielding using the continuum flats yielded no significant improvement and so was not done in all instances. VSpec (astrosurf.com/vdesnoux/) was then used to wavelength calibrate the spectra and subtract the background sky. The spectra from the VATT observing programs were flux calibrated using 14 CVn as a reference star.

The spectra from the earlier Bok telescope observing were either not flux-calibrated or only crudely flux-calibrated using the assumption that the target object had a spectral class of late G or early K.

## 4. Results

Described here are the initial results of over 35,000 photometric observations made over a period of nearly twenty years. Also described are the spectroscopic data obtained to date.

### 4.1. MG1-VSC Observations

Examination of the raw and phased discovery light curve data of MG1-688432 (Figure 1) revealed it to be a periodic variable with sinusoidal variation (P ~ 6.68d), and amplitude of about 0.5 mag.

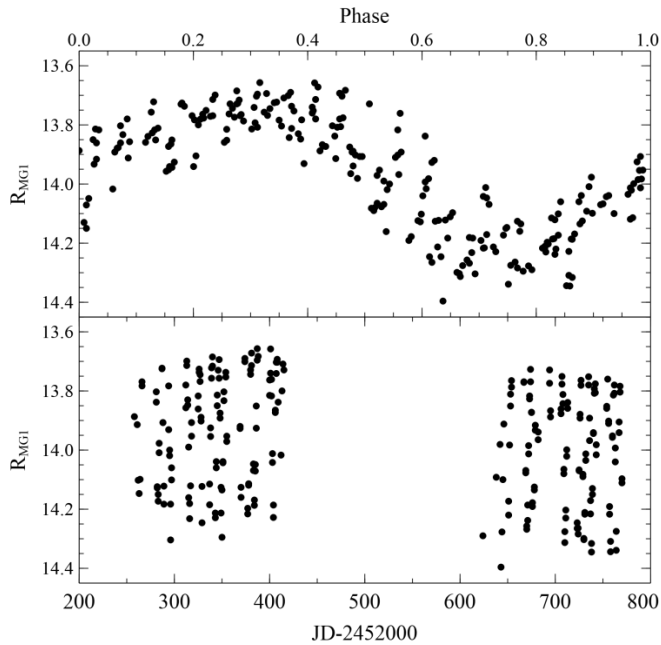

**Figure 1.** This is the MG1-VSC graphical representation of the discovery data for MG1-688432. Upper panel is the phased curve (magnitude vs phase) with period, P = 6.68d; lower panel is the raw light curve.

Basic data for MG1-688432 as extracted from the MG1-VSC are shown in Table 1. These data include the MG1-VSC ID number, equinox 2000.0 right ascension and declination, MG1 R band magnitude, amplitude of the light curve, standard deviation of the amplitude, photometric error, skewness of the magnitude distribution, number (N) of observations, log period (Lomb-Scargle periodogram peak frequency), log of the false alarm probability, and two parameters characterizing the Welch-Stetson statistic. Note that log P = 0.825 corresponds to P = 6.6834 days (a similar period was noted some years later in ASAS data (Jayasinge et al. 2020; Shappee et al. 2014).

**Table 1.**
MG1-688432 basic information from MG1-VSC
(coordinates updated – GAIA)

| Parameter | Value |
|---|---|
| ID(MG1-VSC) | 688432 |
| GAIA DR3 RA(2000) | 12:23:10.16239 |
| GAIA DR3 DEC(2000) | +03:12:37.41873 |
| $R_{MG1}$(mag) | 13.968 |
| Amp(mag) | 0.739 |
| STDEV(mag) | 0.186 |
| Photerr(mag) | 0.181 |
| Skew | 0.59 |
| N | 247 |
| Log Period | 0.825 |
| I_WS | 32.712 |
| sigma_WS | 180.418 |

Stars in MG1-VSC that show sinusoidal light curves as shown in Figure 1 often turn out to be very short period stars with aliased periods arising from the once per sidereal day observing cadence of the MOTESS scan mode programs. As a consequence, the Lomb-Scargle periods determined by the MG1-VSC data pipeline are often in error, which can make these stars desirable targets for quick and easy follow-up observing to determine correct periods. Furthermore, such light curves can be indicative of contact binary systems which may be of interest on their own. These circumstances led to the inclusion of MG1-688432 in a follow-up observing program, as described in Section 4.2. During that program it almost immediately became apparent that MG1-688432 was exhibiting dynamic and unusual behavior that begged more intensive follow-up observation.

## *4.2. Follow-up Photometry*
### *4.2.1. Goodricke-Pigott Observatory*

Shown in this section is a series of graphics (Figure 2) representing phased light curves for each year of follow-up observation (using the twenty year mean period of P = 6.6491 ± 0.0010d). This period is calculated using the combined data set of all photometric observations made over the twenty year observing program. Changes in phase and amplitude were noted during the 20 years, but the mean period shows a tight phased curve for each year of observation. Recall that most of the observations are unfiltered. Observations made in 2015, and shown in this section (Figure 3) are for filtered images only, and there were no observations made in 2016 due to weather. Some figures have specific outbursts marked as Feature F1, F2, etc. to identify specific optical outbursts of interest. These features are described as more general "outbursts" rather than more specific stellar "flares" for two primary reasons: 1) they may have total energies, durations, and morphological shapes different from most stellar flares, and 2) it is possible that their origins are different from "conventional" stellar flares, as discussed later.

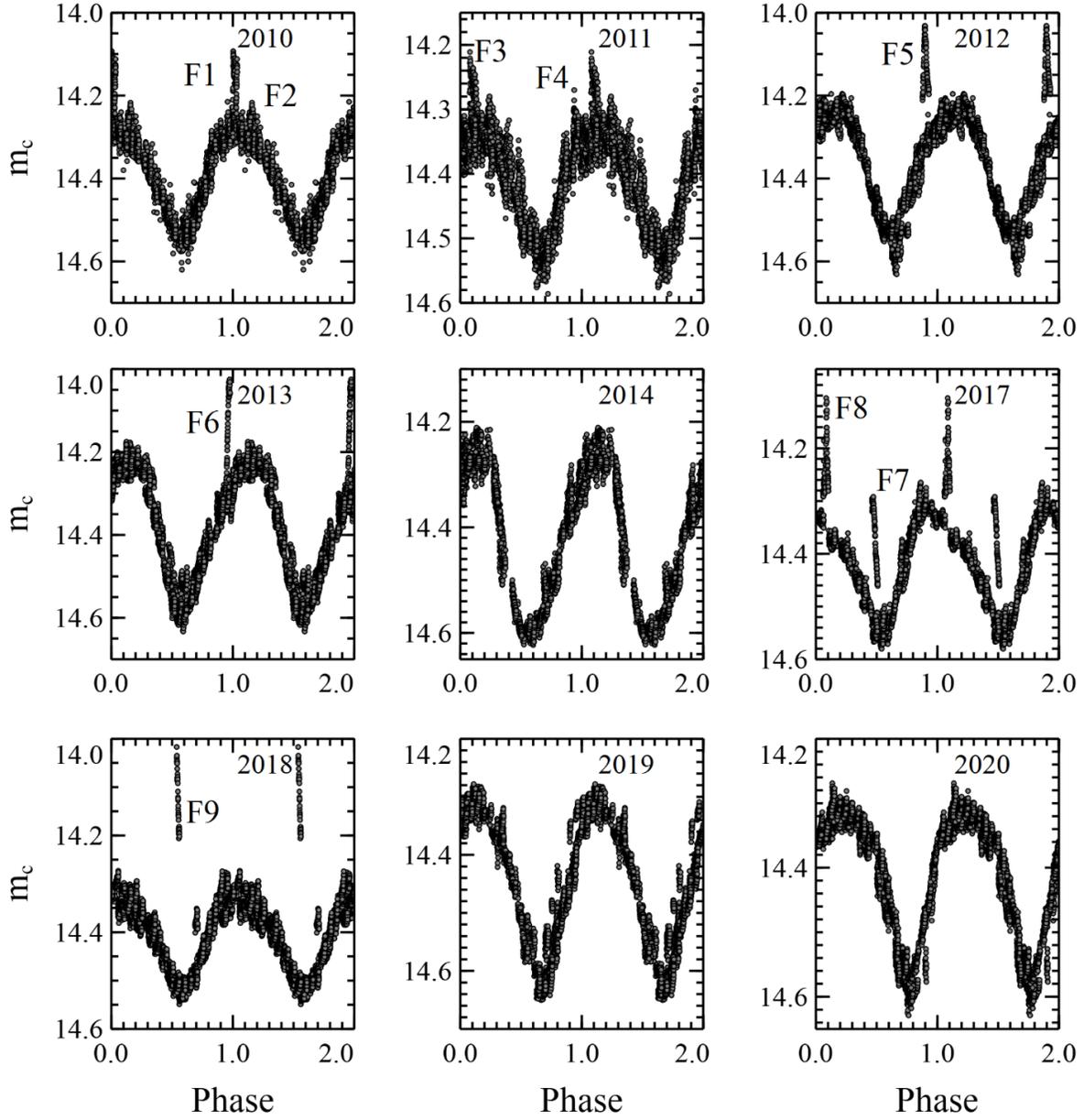

**Figure 2.** GPO follow-up phased light curves based on the 20-year mean period P = 6.6491 ± 0.0010d for MG1-688432.

Observations in 2010 show a quasi-sinusoidal light curve of amplitude ~ 0.3 mag. The first very large outbursts were seen this year. By 2011 the sinusoidal light curve persisted (as it did through the entire twenty years of observation), but the amplitude decreased to ~ 0.2 mag. More large outbursts were observed. During 2012 the amplitude increased to about 0.4 mag. A very bright outburst was observed, and the sinusoidal light curve exhibited distinctly V-shaped minima.

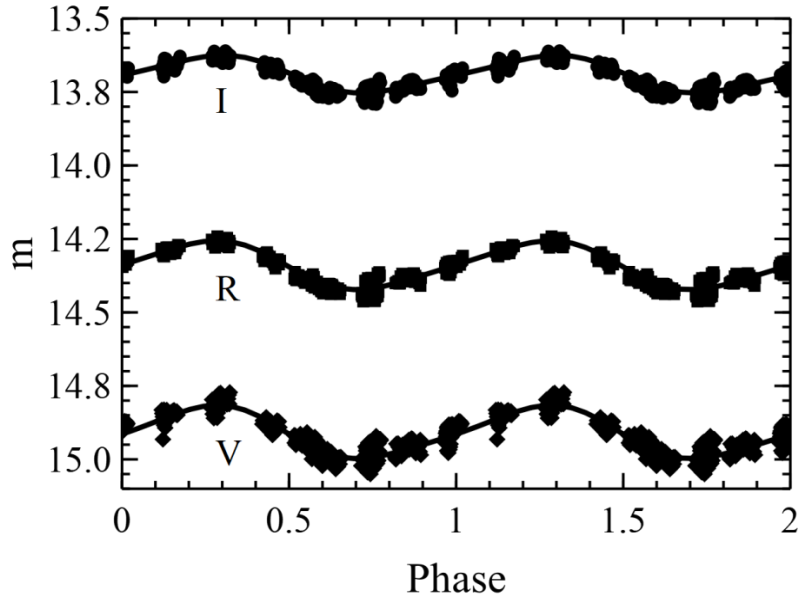

**Figure 3.** MG1-688432 phased light curves (P = 6.6491d) of 2015 in the differential apparent magnitudes in the V, R, and I bands, showing relative brightness and colors.

During 2013 another bright outburst was observed, the V-shaped minima persisted, and the sinusoidal light curve acquired a distinct asymmetry, though the period remained constant. There is a distinct sense of high frequency, low amplitude flickering in the light curve. In 2014 no large outbursts were detected, though the flickering behavior persisted and the light curve asymmetry remained. In 2017 two major outbursts were observed, and the shape of the curve changed dramatically with a steep linear rise and a broken decline in brightness.

The 2018 light curve shows another large amplitude outburst, while the 2019 and 2020 observations did not capture any large outbursts. Both show a continuation of the asymmetry first appearing very distinctly in the 2017 data.

*4.2.2. Follow-up MG1 and MG6 Survey Data*

Using a recently developed GNAT Python data reduction pipeline, the MG1 survey has been re-processed and the MG6 survey (a later repeat of the MG1 survey) has been newly processed. The phased photometric light curves of MG1-688432 as seen in both MG1A and MG6A are shown in Figure 4. Note that MG1A and MG6A indicate the first version databases of the MG1 and MG6 surveys that were considered to have passed first stages of quality control.

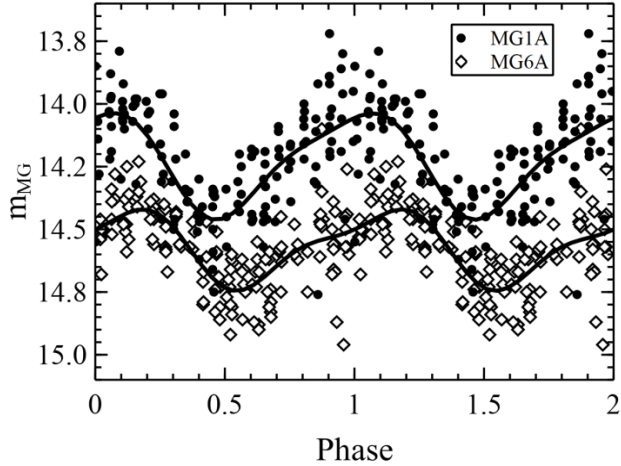

**Figure 4.** Phase diagram for combined data for MG1A (diamonds) and MG6A (circles). The period is 6.64d determined using Peranso. The curves are least squares sinusoidal fits to the data. The ordinate is the MG survey magnitude (open channel) scaled to the SDSS R magnitude.

*4.2.3. Colin Gum Observatory Photometric Data*

Photometric observations of MG1-688432 made at Colin Gum Observatory (CGO) in South Australia are shown in Figure 5. These observations were coordinated with near-simultaneous spectroscopic observations made at the Vatican Advanced Technology Telescope (see Section 4.3).

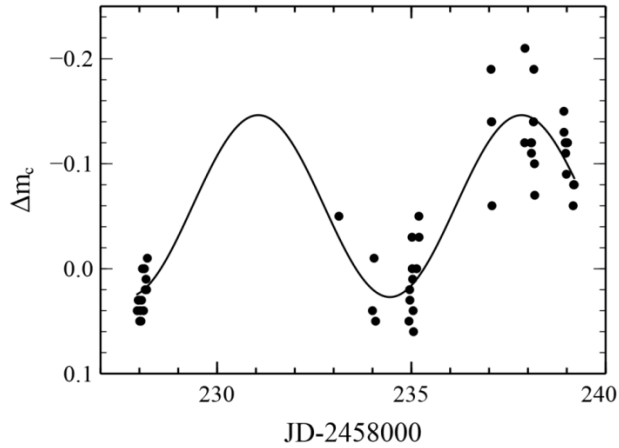

**Figure 5.** CGO observations of MG1-688432 made during 2018. P = 6.8±0.2d.

A sinusoidal model, defined by $M(t) = A_\omega cos(\omega t) + B_\omega sin(\omega t) + C_\omega$ (Cumming et al. 1999), is shown in Figure 5 as the curve fitted to the observations. Least squares estimation of the coefficients $A_f, B_f$ and $C_f$ for a grid of possible circular frequency values, $f$ was carried out which enabled the calculation of a periodogram. The frequency, $f = \omega$, corresponding to the largest peak in the periodogram, defined the optimal parameter set $\{\omega, A_\omega, B_\omega, C_\omega\}$ for the model, $M(t)$. According to the model, minimum light for MG1-688432 occurred at JD 2458234.45, not long after the first VATT spectrum was obtained (see section 4.3), while maximum light occurred at JD 2458237.83, toward the end of the VATT run. The estimated light curve period is 6.8±0.2d.

As the number of observational points is low and the duration of the observing session is also short, it makes sense to calculate an error bar for the period based on a Monte-Carlo method that determines the location of the main peak in the periodogram from many realizations of the data. Each realization is computed by adding a 'fluctuation' to the measured magnitude differential drawn from a Gaussian probability density. The standard deviation in the probability density is the error bar for the individual magnitude differential measurements. The standard deviation of the periodogram peak locations from all the realizations is then interpreted as the error bar, which in the 2018 observations is ±0.2d. The same Monte-Carlo simulation shows that the probability of a "false alarm" is insignificant. Essentially, there is only one main peak in each realization. In situations where there are many more measurements taken over a much longer time, the uncertainty in period or frequency precision would greatly improve. However, the 'false alarm' statistic may become the more relevant quantity to consider.

Further observations of MG1-688432 were carried out for 18 nights in 2019 Jan-Mar. These results are shown in Figure 6.

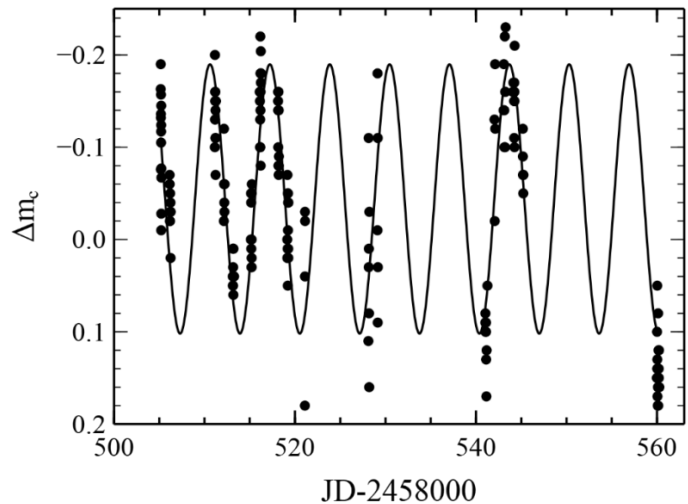

**Figure 6.** CGO observations of MG1-688432 made during 2019. P = 6.61±0.01d.

A light curve model fit to these data yields a period of P = 6.61±0.01d. The first 10 observing sessions were almost consecutive nights and when the model was fit to this data subset a period of 6.55d resulted. These periods may at first appear to be inconsistent but long term observations of MG1-688432 carried out by the Goodricke – Piggott Observatory (GPO) confirm unusual period and phase shift variations over many years. Note that the observations on JD2458528, 2458529 and 2458543 and 2458544 were carried out with a V filter in place. The error bars were determined using the same method as for 2018.

Photometric observations of MG1-688432 coordinated with VATT spectroscopic observations were carried out in 2020 Feb. The light curve obtained from the CGO has a period of 6.62±0.05d and is shown in Figure 7.

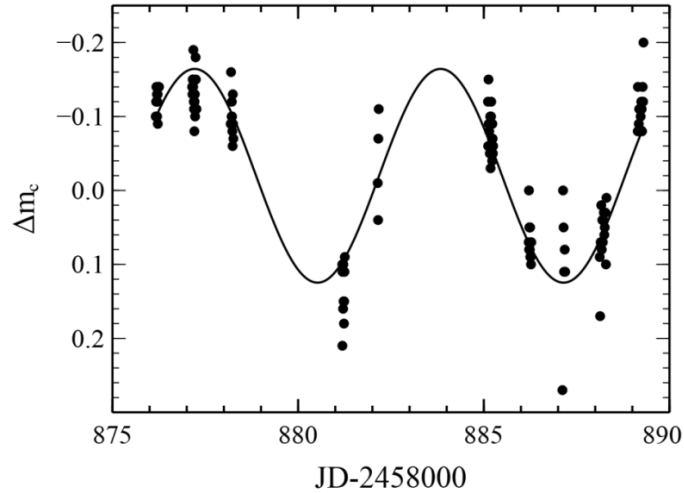

**Figure 7.** CGO observations of MG1-688432 made during 2020. P = 6.62±0.05d.

### *4.3. Spectroscopy*

Spectra of MG1-688432 have been obtained at both the Bok Telescope of Steward Observatory at Kitt Peak, Arizona, and at the Vatican Advanced Technology Telescope (VATT) at Mt. Graham International Observatory, Arizona. An example of the Bok Telescope spectra is shown in Figure 8. VATT spectra were obtained in two overlapping wavelength ranges designed to emphasize either the blue or the red part of the spectrum. An example of each is shown in Figures 8 and 9.

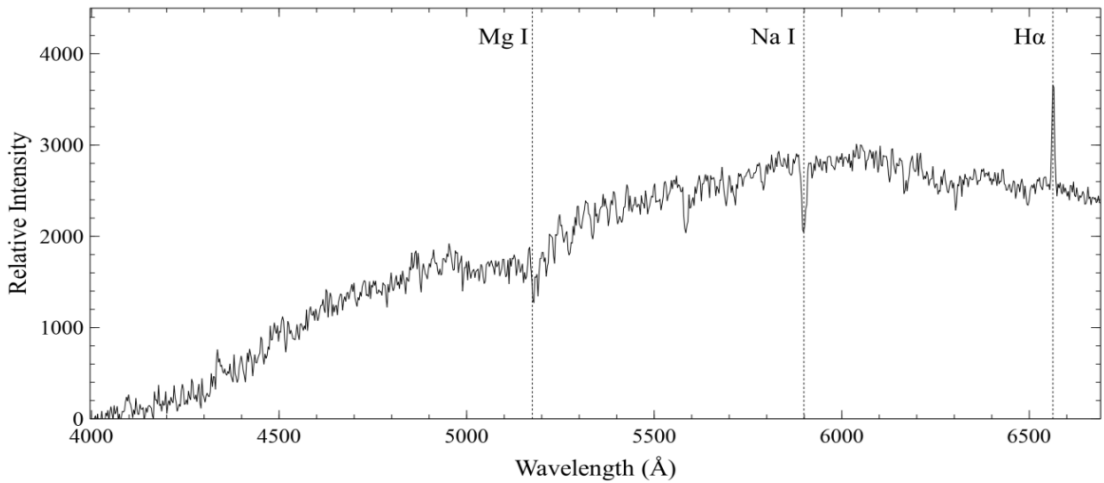

**Figure 8**. A low-dispersion spectrum obtained at the Bok telescope on 2012 Dec 26 showing strong H-alpha emission.

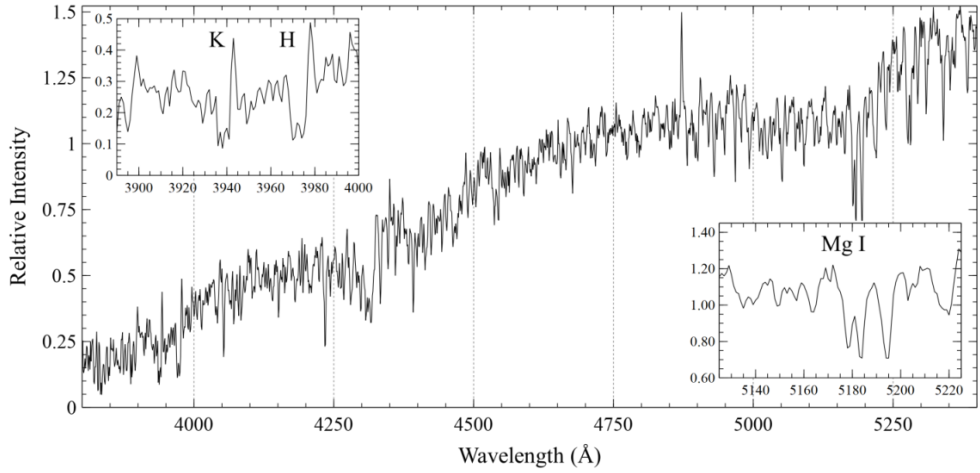

**Figure 9.** A blue spectrum of MG1-688432 obtained at the VATT in 2018. The inserts are of the Calcium II and Magnesium I lines. Note the emission cores in the H and K lines of Calcium.

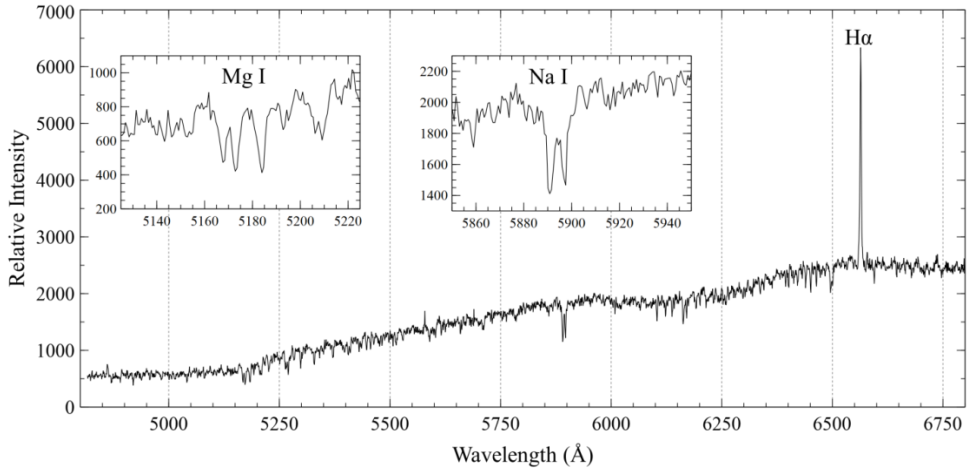

**Figure 10.** A red spectrum of MG1-688432 obtained at the VATT in 2018. The inserts are of the Magnesium I triplet and Sodium D lines.

### 4.4. Serendipitous Observations

In addition to the new observations reported above, MG1-688432 has been serendipitously observed in numerous astronomical surveys, some of which yield information relevant to discussions in this paper. A brief summary of those observations is provided in this section.

#### 4.4.1. Miscellaneous Photometric Data

The Catalina Sky Survey (CSS) is an asteroid search that has archived all of its imaging data for public access (CSS 2020). These images cover much of the sky and include a nine year overlap with MG1 survey and GPO follow-up photometry observations of MG1-688432, as shown in Figures 11 and 12.

The photometric observations provided by the Catalina Sky Survey fill in much of the gap from the end of the original MG1 survey in 2002 to the beginning of the follow-up observations in 2010. The Catalina observations begin in 2005 and overlap the first four years of follow-up through 2013.

The photometry reported in Figures 11 and 12 use different observational systems, but all are sensibly clear channel imaging with similar silicon detectors, allowing for the long term temporal comparisons presented here. It is helpful to note that the CSS and GPO observations overlay one another in the years 2010, 2011, 2012, and 2013.

The observations clearly show a decline in brightness and light curve amplitude until about 2008, at which time the system settles into its current brightness and amplitude range.

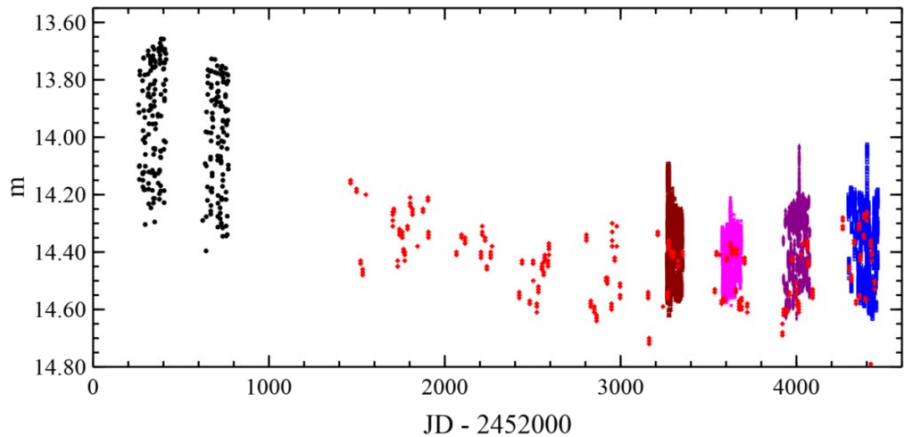

**Figure 11.** MG1-688432 light curves with source photometry as follows: black points: MG1 VSC; red points: Catalina Sky Survey; brown, magenta, purple, and blue are from GPO, 2010, 2011, 2012, and 2013 respectively.

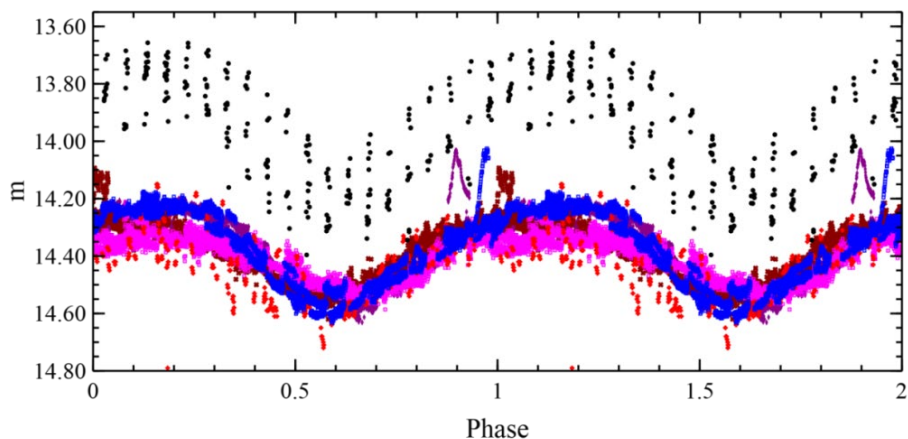

**Figure 12.** This is the same data as Figure 11 except phased with a period of P = 6.6491d, the mean period over twenty years of observation.

The spectral energy distribution for MG1-688432 is represented by a collection of data as shown in Figure 13.

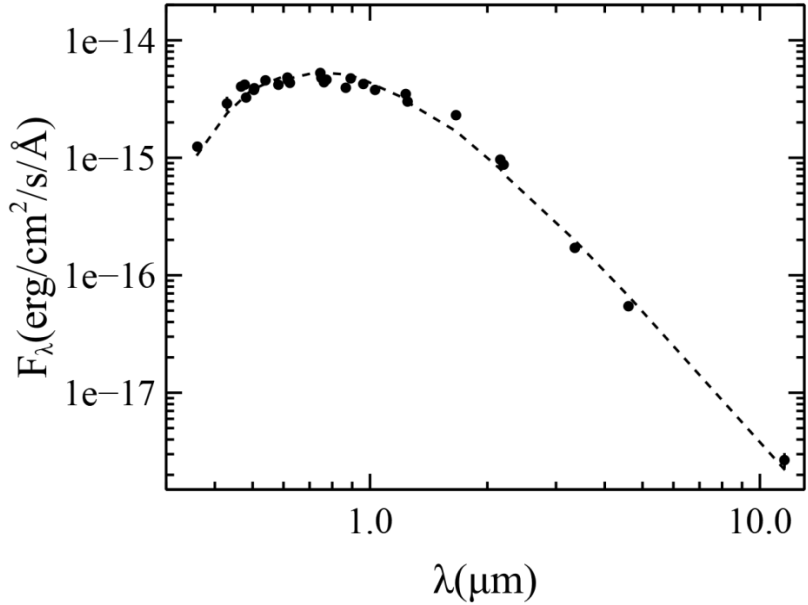

**Figure 13.** Spectral energy distribution for MG1-688432. The black circles are observed flux measurements with error bars from the SDSS, APASS, Pan Starrs, Gaia EDR3, Ukirt, 2Mass and Wise surveys that were collected through the Virtual Observatory SED Analyzer (VOSA) online service (Bayo et al. 2008). The dashed line is a black body fit yielding a Teff of 3950K.

*4.4.2. Gaia Astrometry and LAMOST Spectroscopy*

The Gaia spacecraft (Gaia Collaboration: Prusti et al. 2016) reported observations of MG1-688432 in Gaia Early Data Release 3 (Gaia Collaboration: Brown et al. 2020) as the catalog source Gaia EDR3 3701511568560550400, with significantly nonzero values for both parallax and proper motion. Implications of the Gaia observations are discussed in Section 5.1.2.

The LAMOST stellar parameters pipeline reported a best-fit spectral type of K4 (Luo et al. 2018).

## 5. Analysis and Discussion

*5.1. Discussion of the Observational Data*
*5.1.1. Initial Assessment of MG1-688432 Characteristics*

MG1-688432 (MG1-VSC catalog period P ~ 6.68d) initially attracted attention because of its sinusoidal light curve (suggesting a short-period contact eclipsing binary system) and potentially aliased short period, which could be easily resolved with modest follow-up photometry. Such photometry quickly revealed a period of about 6.65d, or about 13.3d if the star was an eclipsing binary. It also detected an interesting large amplitude outburst that was not expected (see Figures 2 and 18A).

Resolution of the period ambiguity was achieved by noting that the spectra show a single line system. The spectroscopy and associated radial velocity measurements conducted in 2018 and 2020 indicate the presence of one star as part of a binary system with an approximate period of 6.6d. If the longer period were correct (13.2d) there would be spectroscopic evidence of two similar stars, which is not observed. The system comprises two stars (one unseen) that orbit a common barycenter with an inclined orbital plane that does not produce an eclipse. The origin of the sinusoidal light then comes into question, as discussed later.

MG1-688432 was also observed by the Gaia satellite, which measured a distance of 1.5Kpc, implying an absolute magnitude of M = 3.9 mag. Spectroscopy suggested that the star is of about type K3, somewhat evolved off the main-sequence between classes III – IV.

The filtered data obtained at follow-up using B, R, V and I filters was aligned by timestamp for the R, V, and I data to compare data with the smallest time intervals between acquisition. For the V-I calculation the average time lag between acquisition of a V and I paired observation was 16.5 minutes (range of 5.3-123.7 minutes). The main period of MG1-688432 measured using the R magnitude data accumulated contemporaneously was 6.5317d (observations = 494, Peranso ANOVA method). The magnitude of the V-I color was obtained for 458 observations and the period determined to be 6.42d (Peranso, ANOVA method). The Phase diagram for the R and V-I observations (using a Period of 6.5317d) is shown in Figure 14. A convincing sinusoidal curve for the V-I data is observed (Period false alarm probability <0.0001). The V-I color varies in phase with variations in the R magnitude, and the star appears more red when it is fainter (consistent also with the spectra, see Figure 16 and discussion).

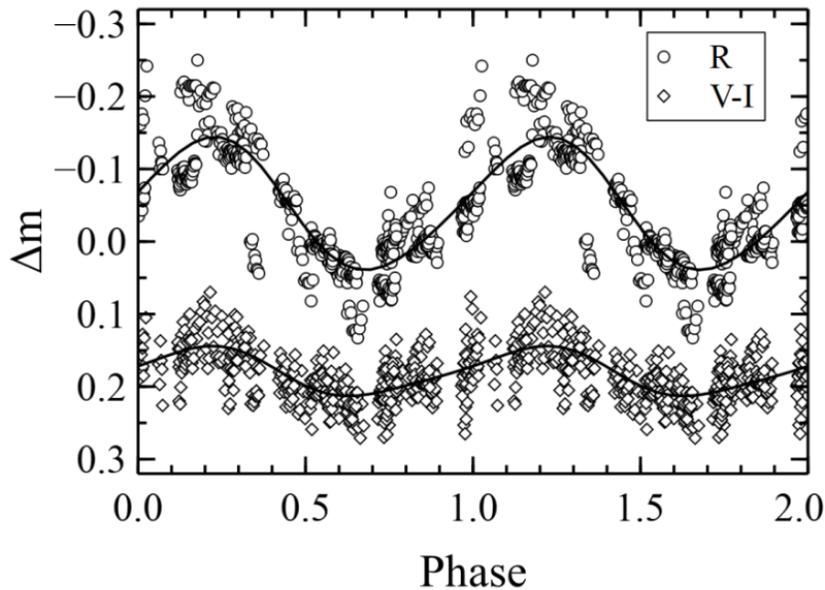

**Figure 14.** Phase diagrams for follow-up R band and V-I color measurements from GPO. The solid curves are least squares sinusoidal fits to the data.

*5.1.2. Astrometry (GAIA Results)*

Given the RA and DEC of MG1-688432 and the Gaia distance of D = 1415 +48/-40 pc (Bailer-Jones et al. 2021) the space position in Sun-centered galactic Cartesian coordinates is (X,Y,Z) = (-163 ± 5, -572 ± 18,

+1284 ± 40) pc. The Z distance is several times the thin-disk scale height of $H_1 \sim 300$ pc, but similar to the thick-disk scale height of $H_2 \sim 900$ pc (Juric et al. 2008).

The Gaia proper motion and distance correspond to a tangential velocity of vtan = 111 ± 3, and adopting a mean systemic radial velocity of 34.7 ± 5 km/s, the inferred space velocity relative to the Sun is (U,V,W) =(33.5 ± 1.3, -111 ± 3.6, -7.1 ± 4.7) km/s (Johnson & Soderblom 1987).

MG1-688432 is located above the Milky Way mid-plane by 4.4 times the thin-disk scale height but only 1.5 times the thick-disk scale height, so the local density of thick disk stars is 2.2 times higher than the density of thin-disk stars (assuming f_thick = 0.12 f_thin at the Milky Way mid-plane; Juric et al. 2008). The observed velocities also place MG1-688432 in a region of the Toomre energy diagram (Sandage & Fouts 1987) occupied primarily by the thick disk stars of the Milky Way and beyond the regime of the thin disk. The thick disk and halo overlap for the position and velocity of MG1-688432, but they have a stellar density ratio of ~5:1 at a height of Z = 1.5 H_thick (Juric et al. 2008), and most halo stars have velocities that are more discrepant from the LSR MG1-688432. The low velocity in the vertical direction also indicates that MG1-688432 recently reached its maximum height above the Milky Way mid-plane, and thus is approximately coplanar with the disk. One therefore concludes that MG1-688432 is more likely to be a thick disk star than a halo star, but confirmation would require a unified chemo-dynamical analysis (e.g., Hawkins et al. 2015).

The Gaia EDR3 catalog entry for MG1-688432 also reports photometry in the broad Gaia G bandpass (spanning 330-1050 nm) and the narrower Bp and Rp bandpasses. MG1-688432 was observed in 38 scans with semi-regular spacing between 2014 July and 2017 May. The reported flux uncertainties in Gaia EDR3 are computed from the RMS if the measured fluxes, divided by the square root of the number of observations, allowing inference that the photometric scatter across this interval was 0.090 mag in G, 0.097 mag in Bp, and 0.067 mag in Rp. MG1-688432 is relatively bright compared to many Gaia sources, so these values of the RMS scatter almost certainly result from the intrinsic variability of the source, rather than photon counting statistics. The amplitudes are consistent with the moderate color dependence of the sinusoidal variability that was seen in our filtered follow-up observations (Section 5.1.1).

Thus one can infer an average absolute magnitude of $M_G$ = 3.69 ± 0.07 mag and an average color of Bp-Rp = 1.430 ± 0.019 mag. In Figure 15 the position of MG1-688432 in the Gaia CMD ($M_G$ versus Bp-Rp) is plotted.

To provide context, the sequences of Gaia Collaboration, Helmi, et al. (2018) are also shown for 14 globular clusters spanning a range of metallicity, as well as the sequences from Gaia Collaboration, Babusiaux, et al. (2018) for the old solar-metallicity open clusters NGC 188 (T = 5.5 Gyr) and NGC 2682 (T = 3.5 Gyr). Intriguingly, MG1-688432 sits ~5 magnitudes above the main sequence, which implies that there is at least one evolved star in the system. However, it also sits red-ward of the red giant branch for even the most metal-rich old cluster. This regime of parameter space is sparsely occupied by stars that have undergone non-standard stellar evolution due to binary interactions (e.g., Belloni et al. 1998; Mathieu et al. 2003), and MG1-688432 falls near the boundary between "sub-subgiants" and "red stragglers" (Geller et al. 2017).

However, a star beginning to ascend the red giant branch could also be reddened to this position either by interstellar dust or by local dust within the system.

Figure 15 also shows the reddening vector in (A(G), E(Bp-Rp)) that corresponds to an extinction of A(V)=1.0 mag, using the Gaia reddening relations of Danielski et al. (2018) and Wang and Chen (2019). If MG1-688432 were a normal giant, the required extinction for a standard interstellar reddening law would be A(V) > 1.0 mag, depending on its age and metallicity. This extinction cannot be explained by interstellar extinction; the Bayestar19 dust maps of Green et al. (2018) report a line-of-sight extinction of A(V)=0.0 mag along the line of sight to MG1-688432, integrated entirely out of the Milky Way, which is not unexpected for a star that is located <30 degrees from the North Galactic Pole. One can therefore conclude that the HR diagram position of MG1-688432 is established by an intrinsic feature of the system, most likely either the stellar evolutionary state of the observed star or the presence of small (non-gray) dust within the system.

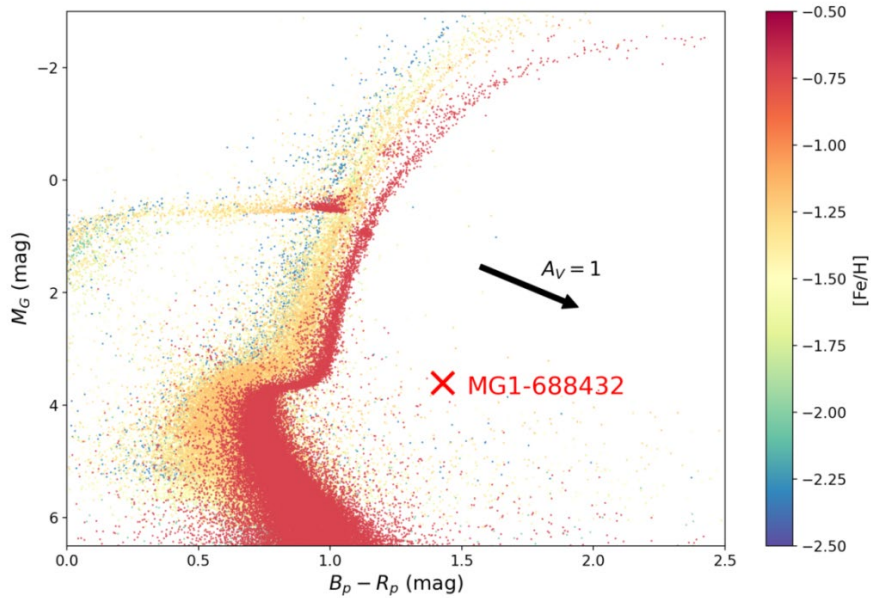

**Figure 15.** The position of MG1-688432 is shown in the Gaia color magnitude diagram (MG vs Bp – Rp), as well as the cluster sequences of Gaia Collaboration, Helmi et al. (2018).

*5.1.3. Spectroscopy and Radial Velocities*
5.1.3.1. Analysis of the VATT Spectra

Figure 16 shows a montage with VATT spectra of MG1-688432 from four nights in 2018 Apr. A MK standard star spectrum is placed at the top (alp Ari = K1 IIIb) and bottom (HR 753 = K3 V) for comparison. These standard spectra were taken at VATT on another occasion with a 1 arcsec slit but with the same 600 g/mm grating as used for MG1-688432. They were re-binned and smoothed to have a resolution equivalent to the 1.5 arcsec slit used for MG1-688432.

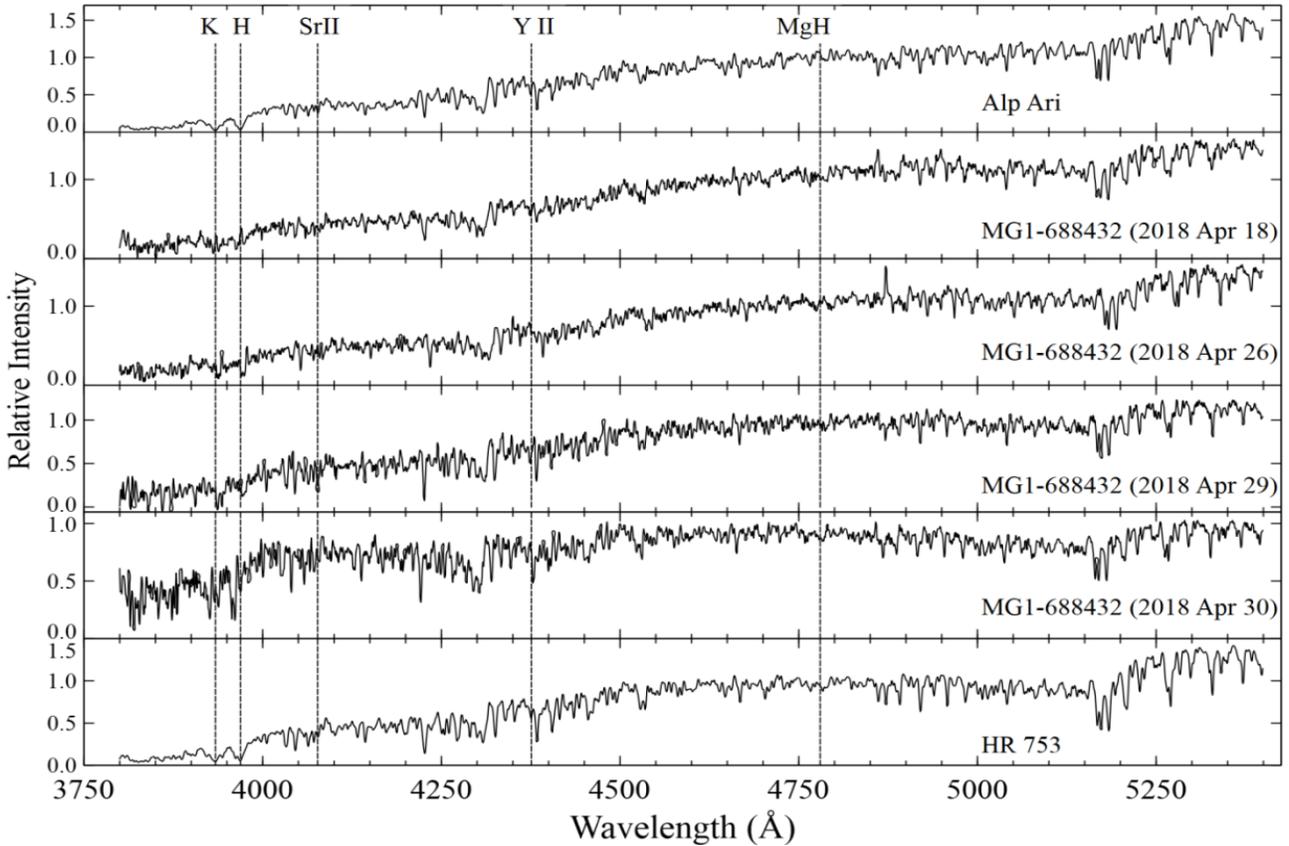

**Figure 16.** This is a spectral montage, with the four blue-region spectra of MG1-688432 from 2018 Apr, and with MK standards alp Ari (K1 IIIb) on top and HR 753 (K3 V) on bottom. These spectra are normalized and offset.

This procedure was checked for the G8 III standard, kap Gem, observed in 2020 Feb in the blue region at the same configuration as MG1-688432. Another G8 III standard, omi Psc, taken with the 600 g/mm grating, was rebinned and smoothed like alp Ari and HR 753. The two G8 III spectra at their common resolution matched ratios closely, though the continua and depths were different at each end since omi Psc was not flux corrected.

The four blue-region spectra of MG1-688432, which are clearly not quite the same, average to an MK type of K3 III-IV:e: CN-3 ke P Cyg.

Given the varying emission in the hydrogen lines, the temperature type is estimated from ratios of the chromium I ($\lambda\lambda 4254, 4275, 4290$) to nearby iron I lines. These showed a fairly consistent K3. Luminosity is given by the ratios of Sr II $\lambda 4077$ and $\lambda 4216$ to nearby Fe lines, by Y II $\lambda 4376$ similarly, and in mid K-type stars by MgH around $\lambda 4780$. It varied between III and V, with III-IV as the average. The CN band short of Sr II $\lambda 4216$ would normally be used for luminosity also, but no significant CN is apparent. Since the G-band is normal, nitrogen is presumed quite weak.

The spectra have a signal-to-noise between 30 and 50. All are somewhat noisy, but there are considerable inconsistencies of the line depths from spectrum to spectrum, which are more than would be due to their S/N. None match a K3 IV standard well, even allowing for the emission, so MG1-688432 is variably peculiar.

Starspots usually do not produce such large variations and "mismatches" with an MK standard. For instance, HD 222107 (λ And) is an RS CVn binary whose primary has variable Ca II K & K emission from at least two large starspots (Morris et al 2019), and yet its spectrum matches a G8 IV MK standard well (Gray et al. 2003) save for the emission. However, absorption features that change rapidly around a particular temperature class can betray the presence of cooler regions in starspots (see TiO band modeling of Morris et al 2019). There is some evidence in Figure 16 for MG1-688432 to be showing slightly more absorption from TiO at λ4955 on 2018 Apr 18, when MgH at λ4780 could also be stronger, and there seems a trace of TiO also on 2018 Apr 30.

The hydrogen lines show varying emission, which is particularly obvious in H-beta. Chromospheric emission is evident in the Ca II H & K lines (the "ke" index in the classification), though there is much noise this far into the blue. Three of the four spectra show in Ca II H & K the unmistakable P-Cygni profile, again in varying strengths that follow the hydrogen emission strength. An expanding gas and dust envelope, variously excited, could well cause these variations and general inconsistencies.

In agreement with the end of Section 5.1.2, no persistent reddening, due to interstellar extinction, is apparent in MG1-688432's flux calibrated spectra (Figure 16). Variable dust within the system would not cause the small spectrum-to-spectrum variations in line depths, though something, perhaps instrumental, has lowered the SED toward the red on 2018 Apr 30.

If MG1-688432 is a normal giant star near the base of the giant branch, it should have an effective temperature of 4800 K < T_eff < 5200 K (e.g., Hidalgo et al. 2018), corresponding to a spectral type no later than ~K0III (e.g., Gray and Corbally 2009, Appendix B). If the observed broadband colors are instead indicative of the intrinsic photospheric properties, then MG1-688432 should have a spectral type of K3III-K4III (Ramirez & Melendez 2006). While extinction could redden the broadband colors, the system should still exhibit the spectral absorption features of its intrinsic effective temperature, and thus the spectral type can test whether there was substantial extinction present during any of the spectroscopic observational epochs. The discussion of MG1-688432's spectral type shows that it is comfortably later than K0 III-IV, despite the peculiarities, and so has an effective temperature of around that indicated by the SED of Figure 13, namely 3950K.

Red region spectra were taken with the same 600 g/mm grating and with a range of 4830–6800 Å on three nights, 2018 Apr 26, 28, and 29. They are consistent with the blue classifications in temperature type and varying emission strength.

The Bok 90" spectra from 2012-13 are not flux calibrated and range from 3700–6900 Å. Their resolution and profiles match quite well in resolution with the Dark Sky Observatory (Appalachian State University) 3.6A spectra of MK standards that range 3800–5600 Å. They have an average classification of K2: IV:e ke. All have H-alpha in emission. Again, like VATT spectra, the ratios and line depths are very variable, but indicate an early-K, slightly evolved star.

In conclusion, MG1-688432 is not like any normal star, single or composite. "K3 III-IV:e: CN-3 ke P Cyg" is an approximate description, as the colons indicate, since it varies night-to-night and never matches an MK standard particularly well.

5.1.3.2. Radial Velocity Measurements

Two methods for determining the radial velocity variations of MG1-688432 were used in the analysis. The first method relied on the accurate modeling and position determinations of individual spectral lines. The second method relied on cross-correlation of broad portions of the spectrum potentially containing many lines.

*Individual spectral line positioning* (Method A): In preparation, every spectrum was first flat fielded, and then wavelength calibrated. Individual integrations were typically of 15 minutes duration and several (typically 3-4) frames were median combined to eliminate high frequency Shot noise events. After sky subtraction, several (normally between 6 and 8) star spectra were averaged per night to yield a spectrum ready for radial velocity processing. The exception was the night of 2020 February 3 when only one spectrum was recorded.

Consistent results using the line centroiding function of the Vspec analysis software were obtained from five spectral lines at rest frequencies 5183.60 Å (Mg I), 5328 Å (blend of 5328.038 Fe I, 5328.323 Cr I, and 5328.531 Fe I), 6102.73 Å (Ca I), 6122.23 Å (Ca I), and 6162.18 Å (Ca I). The radial velocities from each line were averaged to obtain one radial velocity reading and the standard deviation of the radial velocity estimates served as an uncertainty measure. The results of the analysis are summarized in columns 2 and 3 of Table 2.

*Spectrum cross-correlation* (Method B): The averaged spectra described in Method A were cross-correlated with the average spectrum from 2020 February 7 acting as the "standard spectrum" in each cross-correlation. This spectrum produced the most accurate velocity determination using Method A.

Method B potentially yields radial velocities with uncertainties of the order of a kilometer per second and was applied to two 300 Å wide regions. The first region was centered on Magnesium triplet which also contains less prominent Calcium and Iron lines, the second on the sodium doublet. The radial velocities from method A were augmented with the method B velocities to produce more accurate estimates.

The cross-correlation process used in this analysis was similar to the method outlined in da Costa et al. (1977). Furthermore, uncertainties in the radial velocity estimates were calculated using a scheme formulated by Tonry and Davis (1979).

**Table 2**
Heliocentric radial velocities, (RV) and uncertainties, ($\Delta$)

| Date | Method A | | Method B | | | | Weighted average | |
|---|---|---|---|---|---|---|---|---|
| | | | Mg triplet (5000Å-5300Å) | | Na doublet (5750Å-6050Å) | | | |
| 2020 | RV(km/s) | $\Delta$(km/s) | RV(km/s) | $\Delta$(km/s) | RV(km/s) | $\Delta$(km/s) | RV(km/s) | $\Delta$(km/s) |
| Feb 3 | 48.2 | ±20 | 50.6 | ±10 | 44.0 | ±8 | 46.7 | ±6 |
| Feb 5 | 5.5 | ±15 | 12.3 | ±9 | -0.8 | ±6 | 3.4 | ±5 |
| Feb 6 | 23.6 | ±11 | 18.2 | ±9 | 17.9 | ±5 | 18.7 | ±4 |
| Feb 7 | 44.7 | ±10 | 44.7 | ±9 | 44.7 | ±5 | 44.7 | ±4 |
| Feb 8 | 55.1 | ±10 | 65.2 | ±8 | 57.4 | ±6 | 59.2 | ±4 |

The heliocentric radial velocity results and associated uncertainties from the Mg triplet and N doublet spectral regions are given in Table 2. The velocities are close to those obtained from method A and all values lie within each other's error bars. A weighted average of all the radial velocity results and associated uncertainties is given in the last two columns of Table 2. Augmenting the Method A and Method B results yields uncertainties between ±4 km/s and ±6 km/s which is a significant improvement over method A alone.

Spectroscopy undertaken at the VATT during 2020 Feb shows time varying Doppler shifts consistent with a binary system. However, the absence of any line broadening, splitting, or the periodic appearance of lines is compelling evidence that either the system is not an eclipsing binary, or that one of the stars was too faint to be spectroscopically detected. Preliminary radial velocity modeling based on the classical Newtonian Two-Body problem shows that the radial velocity time series resembles one stellar component moving as part of a binary star system.

For the special case of circular orbits, the radial velocity is a sinusoidal function of time and as depicted by the solid curve in Figure 17, a sine function fits the data well. Furthermore, the amplitude of the sinusoid is dependent on the sine of the inclination angle, as well as the masses of both stellar components and the orbital period. If the inclination angle is set to 90°, suitable choices of stellar masses can be made that produce the same radial velocity curve in Figure 17. In such a scenario, the distances between the two stars would infer that the binary system is detached, and if one of the stars is invisible, the observed light curve would resemble a "flattish" line with periodic dips. This is not what is observed. The shape of the light curve further supports the notion that the system is not an eclipsing binary. On the other hand, if assumptions are first made about the possible stellar masses, the resulting inclination angles are too low for eclipsing situations. Further evidence supporting the rejection of an eclipsing scenario comes from the phase shift in the light curve relative to radial velocity curve in Figure 17. Maximum and minimum light is observed when the binary components are well separated and not along the line of sight.

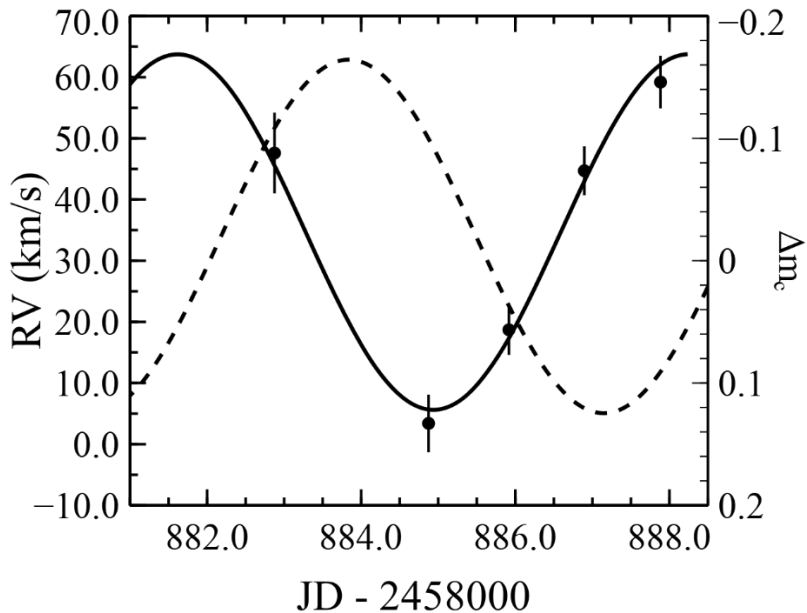

**Figure 17.** The weight averaged radial velocity curve for MG1-688432 derived from the 2020 VATT spectroscopic observations. The period of the radial velocity curve and the overlaid (dashed) light curve is 6.62d.

A detailed parametric analysis of the orbital elements which discusses constraints on the stellar masses and the nature of MG1-688432 is the subject of a future paper. For now it is sufficient to say that the MG1-688432 is most likely a binary star system and the sinusoidal nature of the temporal variation in radial velocity suggests circular or near circular orbits with a systemic velocity of around 34.7 km/s. Spectroscopy also makes it clear that MG1-688432 is not an eclipsing binary, since an equal-brightness pair of near-contact stars would have been detectable as an SB2 in the spectra.

### 5.1.4. Characteristics of the Outbursts

The APASS magnitude of MG1-688432 is V~14.78 (APASS 2020) at a GAIA distance of 1.5 kpc. The luminosity of MG1-688432 is therefore a minimum of about $9 \times 10^{33}$ erg/s. Nine large outbursts have been observed as shown in Figure 18 A-I.

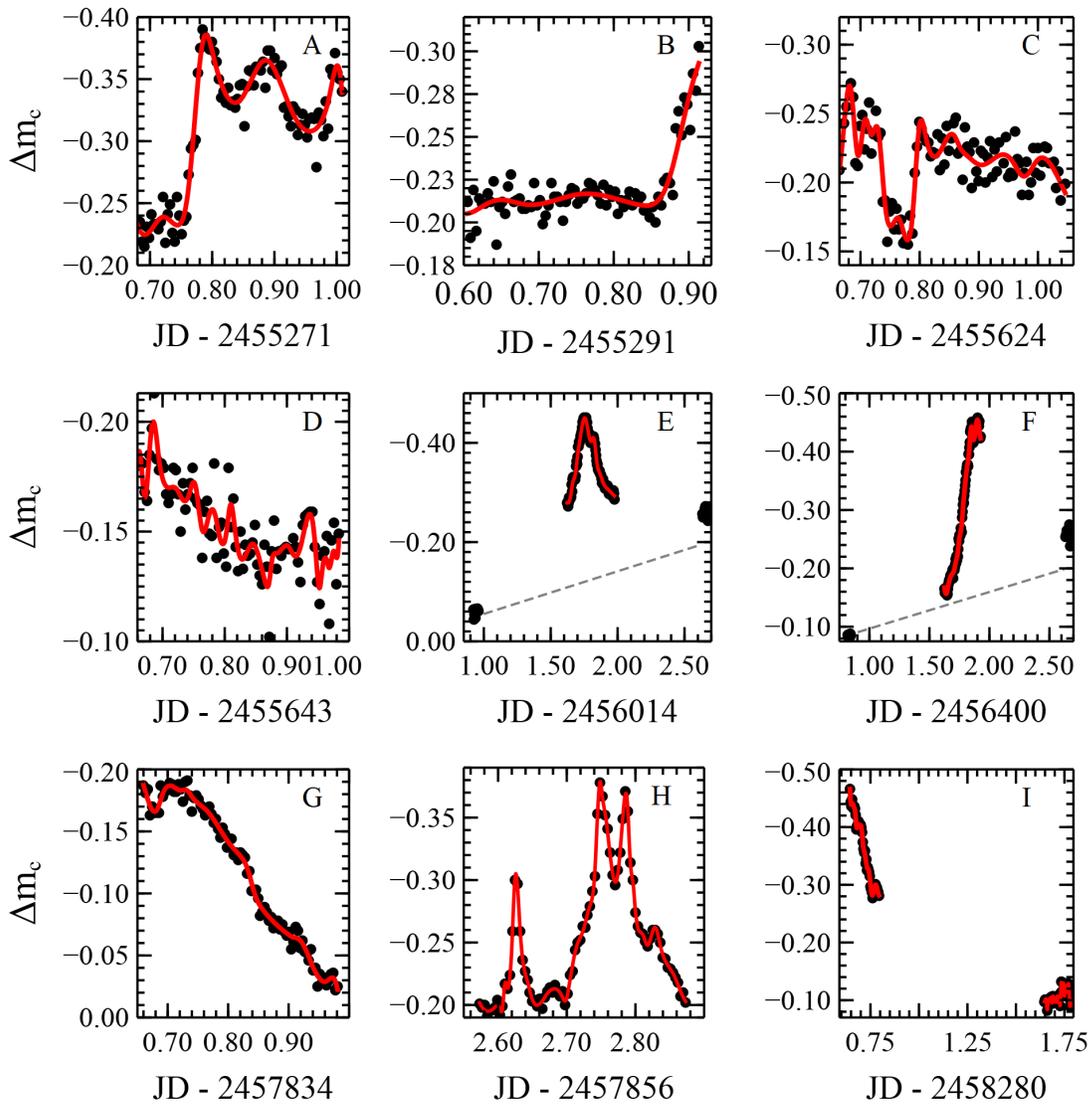

**Figure 18 (A – I).** The nine large outbursts observed in MG1-688432 (compare to the light curves of Figure 2.

The first observed outburst, which serves as a zero phase reference event, occurred on JD 2455271.8 (Figure 18A). The average brightness of the event was 0.09 magnitudes for >21,000 seconds. Observations ended by morning twilight, thus yielding a minimum total energy of $>1.6 \times 10^{37}$ ergs.

The second observed outburst was observed about JD 2455291.9 at phase 0.028 (Figure 18B). The beginning of the outburst started 4,000 seconds before morning twilight interrupted. The light was steeply increasing at the end of observations. The average power for the observed period was $3.4 \times 10^{32}$ ergs/s, yielding a minimum total energy of $> 1.36 \times 10^{36}$ ergs.

The third major outburst observed was at JD 2455624.7, phase 0.082 (Figure 18C). The object appears to be in eruption all of the observable night except for a remarkable interruption of a little over one hour after about one hour into the evening. There is a slow downward trend in brightness from about 0.07 magnitudes above expected to about 0.04 magnitudes for an average magnitude of about 0.05 mag. This leads to an average power of $4.7 \times 10^{32}$ ergs/s and a total energy (allowing for the one hour interruption) of $6.6 \times 10^{36}$ ergs.

Outburst #4 was caught on JD 2455643.7, at a phase 0.94, as just the decay from a higher brightness level as the observing period began (Figure 18D). Observation started at 0.05 magnitudes above the expected level and decayed to zero in about 14,000 seconds. Average power during the period was 0.025 magnitudes or $2.1 \times 10^{32}$ ergs/s. Total energy in 14,000 seconds was $2.9 \times 10^{36}$ ergs.

Outburst #5 was seen on JD 2456015.6 at phase 0.87 (Figure 18E). A very nicely observed event caught on the rise at the beginning of the observing period reached a peak in the middle of the night, and decayed during the remainder of the night. Average brightness above normal levels was about 0.18 magnitudes or $1.6 \times 10^{32}$ ergs/s. The observed duration was 30,500 seconds for a total energy of about $4.9 \times 10^{37}$ ergs.

Outburst #6 was observed on JD 2456401.7 at phase 0.94 (Figure 18F). This outburst was caught on the rise and plateaued in the middle of the night and stayed there for the remainder of the observing period. Some smaller activity was seen the nights before and after. Peak brightness was about 0.25 magnitudes above the normal level or $2.3 \times 10^{32}$ ergs/s. Integrated over 11,400 seconds, this corresponds to a total observed energy of $2.6 \times 10^{37}$ ergs.

Outburst #7 was observed on JD 2457834.7 at phase 0.468 (Figure 18G). This outburst is opposite of the 2013 event. The observing period begins with the light at a plateau of 0.22 magnitudes above the expected level and then decays to a much lower level by the end of the night. 0.22 magnitudes corresponds to approximately $2.0 \times 10^{32}$ ergs/s and integrated over 18,000 seconds yields a minimum energy of $3.6 \times 10^{37}$ ergs.

Outburst #8 was seen on JD 2457858.6 at phase 0.06 (Figure 18H). This was an amazing outburst! It was in eruption all night at an average power of 0.1 magnitudes or $8.7 \times 10^{32}$ ergs/s. However, there are three sharp spikes in brightness of about 0.1 magnitude peak, each lasting about 30-40 minutes. Duration of the night was 26,000 seconds for a total integrated energy of $2.2 \times 10^{37}$ ergs.

Outburst #9 was observed on JD 2458280.7 at phase 0.55 (Figure 18I). This is the largest outburst observed in this star to date. As observations began at evening twilight the brightness was rapidly declining from 0.55 magnitudes above normal level. After slightly over three hours, the brightness was still 0.3 magnitudes above normal. The next night, it was still about 0.1 magnitudes above normal for the over three hour observing period with 0.02 magnitude outbursts super-imposed; see also Figure 9. It is likely that activity continued from the first to the second night and it is possible to interpolate the energy during that period.

Average brightness the first night of 0.4 magnitudes corresponds to $4.0 \times 10^{33}$ ergs/s (greater than solar luminosity) × 10,800 seconds = $4.3 \times 10^{37}$ ergs. Estimated average brightness during the unobserved period was 0.2 magnitudes. This corresponds to $1.8 \times 10^{33}$ ergs/s × 72,000 seconds = $1.3 \times 10^{38}$ ergs. Average brightness the second night was about 0.1 magnitudes, which corresponds to $8.7 \times 10^{32}$ ergs/s × 10,800 seconds = $9.4 \times 10^{36}$ ergs. Total energy is estimated to have been $1.8 \times 10^{38}$ ergs.

Most outbursts are about $1.5 \times 10^{37}$ ergs and last hours, sometimes with considerable structure. The largest observed outburst (#9) produced a minimum of $1.8 \times 10^{38}$ ergs in a period of over a day. A major solar flare lasts minutes and produces $\sim 10^{32}$ ergs. Stellar flare energies can exceed $10^{37}$ ergs.

Figure 19 provides a summary of the outburst events as a function of phase using the mean period, P = 6.6491d. The strong visual impression is that the outbursts are clumped in two groups separated by about 180°, though null hypothesis testing suggests that, due to the small number statistics, this is not a statistically significant correlation.

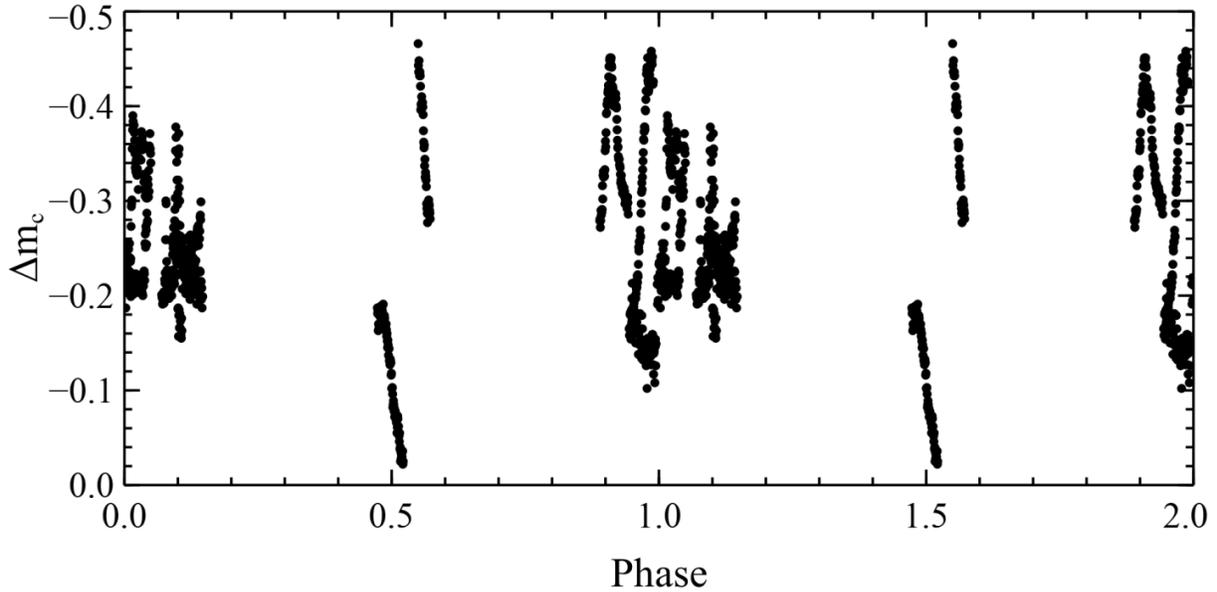

**Figure 19.** Raw light curve data for the outbursts using P = 6.6491d suggesting the two nodes of outburst activity tied to the phase of the star's brightness.

### 5.2. Possible Models for MG1-688432

Radial velocity measurements confirm that the object is at least a binary system orbiting a barycenter However, there is no direct observation of the companion star in the spectra, which do not exhibit double lines, indicating that the companion is much fainter than the primary star. This in turn confirms that the observed light curve arises from the primary, that the ~6.6d period is correct for the system, and that this is not an eclipsing system.

A complication for developing a satisfactory explanation is that the source of the outbursts has not yet been determined. Do the eruptions originate on the K3 component or the unseen companion? Two hypotheses that could be invoked to explain the behavior of MG1-688432 are indicated below. Both of these models, or a combination of the two, will be the subject of future research.

The K3 component is a chromospherically-active star. It may be that the observed outbursts are extraordinarily powerful stellar flares due to strong convection in the stellar envelope (Karoff et al. 2016; Kovari et al. 2020). In this case, the unseen component may simply be a red dwarf. The quasi-sinusoidal light variations may be explained as persistent, large-scale starspots. It is, however, somewhat troubling that the spectra are not wholly consistent with this conclusion, and that the outbursts seem different in character from normal stellar flares.

A second possibility is that the unseen companion in this highly evolved system is a high gravitational potential object, most likely a white dwarf star. In the recent past, a free-floating planetary body may have passed within the Roche limit of the white dwarf and was disrupted into a large amount of fragmentary debris (cf. Gansicke et al. 2016; Malamud & Perets 2020). Interaction with the remnant curtain of material might then contribute to both the sinusoidal variation (passage of the primary through an obscuring debris cloud) and the outbursts (debris impacts on the white dwarf companion).

The second, admittedly exotic, model is of interest for four compelling reasons: 1) the spectra are not completely consistent with the magnetic activity model, 2) the outbursts appear to be distributed by phase consistent with orbital plane crossing events, 3) the outbursts have unusually large total energies consistent with impacts of planetary debris sized objects, and 4) the outbursts bear little resemblance in shape to conventional stellar flares (Balona 2015; Davenport 2015; Gunther et al. 2020). For these reasons at least, the impact model, as well as the more conventional magnetic model, are deserving of future study.

## 6. Conclusions and Future Observations

Although the light curve has a period of about 6.65d averaged over the twenty years of observations, there are unusual amplitude, phase and shape variations that are seen from year to year. The light curve has been observed to vary from a sinusoidal to a skewed sinusoidal, to a "bouncing ball" shape with amplitude varying between 0.2 to 0.4 m. Between 2010 and 2018, nine high energy outbursts have been observed, the largest being at least $1.8 \times 10^{38}$ ergs. Spectroscopy shows the presence of one K3 sub-giant star which also exhibits strong H$\alpha$ and Ca lines, typical of a chromospherically active star.

Radial velocity measurements infer that the star is a member of a binary system; the other member being invisible. However, preliminary radial velocity modeling, as well as the spectroscopy, infer that the system is not an eclipsing binary. Speculation surrounds the nature of the unseen component, but lack of observed X-ray emissions from the region would make it unlikely that the invisible star is a black hole or neutron star.

The nature of the high energy flare events remains a mystery. To explain the various observations, two hypotheses have been proposed which require both theoretical and modeling work combined with further observations. The first speculates that the visible star exhibits magnetically induced chromospheric activity and that the periodic variations exhibited by the photometry are due to star spots. In this scenario, the high energy outbursts would be due to unprecedented flaring events. The second speculates that the high energy outbursts were due to the disruption of a planetary body (likely a free-floating planet given that age and evolutionary state of the star) by the unseen component.

It is clear that this star is deserving of further observation as briefly listed here.

Large telescope, high resolution spectroscopy could be useful in an attempt to learn more about the presently unseen companion star in MG1-688432. Additional spectroscopy at a range of phases may also be useful in helping to understand some of the spectroscopic anomalies observed in the K3 star. In particular, investigation of the strength of its magnetic field should be possible through Zeeman splitting and changes in molecular features (Afram & Berdyugina 2015).

Additional spectroscopy can also confirm details of the orbit (circular or eccentric) and the mass of the companion, which will help in understanding how interactions between the two stars may be affecting the primary. If there are tidal interactions or mass transfer onto a compact stellar remnant, that may also be detectible.

A continuation, and longitudinal expansion, of current time series photometry programs will be very useful in characterizing the photometric evolution of the star. This will help address questions of the temporal frequency variation of the outbursts, as well as the evolution of their magnitudes.

Time series photopolarimetry should be able to record the modulation of polarization as the K3 star orbits its barycenter and periodically encounters the hypothesized debris plane.

If the magnetically induced chromospheric model is correct, leading somehow to extraordinarily energetic outbursts, then one might expect to see significant emission in both x-ray and radio frequencies. The amplitude of these signals may also be temporally sensitive to the orbital period of the system, or may only be observed during optical outburst episodes.

### Acknowledgements

The authors express sincere appreciation and thanks to Dr. Elizabeth Green of the University of Arizona's Steward Observatory. She kindly provided an hour of her observing time with the Bok 90" telescope to acquire the first spectrum of this star and she shared abundant advice and encouragement.

The authors appreciate grants of time to conduct spectroscopy with the excellent facilities of the Vatican Advanced Technology Telescope (VATT) of the Vatican Observatory (VO) and Steward Observatory. They especially appreciate the friendly and generous help provided by VATT personnel, and their patience in training observers new to this facility.

This research benefitted from archival data provided by several sources as acknowledged below.

APASS: The AAVSO Photometric All-Sky Survey (APASS), funded by the Robert Martin Ayers Sciences Fund and NSF AST-1412587.

CSS: The Catalina Sky Survey is funded by the National Aeronautics and Space Administration under Grant No. NNG05GF22G issued through the Science Mission Directorate Near-Earth Objects Observations Program. The CRTS survey is supported by the U.S. National Science Foundation under grants AST-0909182.

GAIA: This work has made use of data from the European Space Agency (ESA) mission Gaia (https://www.cosmos.esa.int/Gaia), processed by the Gaia Data Processing and Analysis Consortium (DPAC, https://www.cosmos.esa.int/web/Gaia/dpac/consortium). Funding for the DPAC has been provided by national institutions, in particular the institutions participating in the Gaia Multilateral Agreement

LAMOST DR4: Guoshoujing Telescope (the Large Sky Area Multi-Object Fiber Spectroscopic Telescope LAMOST) is a National Major Scientific Project built by the Chinese Academy of Sciences. Funding for the project has been provided by the National Development and Reform Commission. LAMOST is operated and managed by the National Astronomical Observatories, Chinese Academy of Sciences.

VizieR: This research has made use of the VizieR catalogue access tool, CDS, Strasbourg, France (DOI : 10.26093/cds/vizier). The original description of the VizieR service was published in 2000, A&AS 143, 23.

**ORCID iDs**

Christopher J. Corbally  https://orcid.org/0000-0001-6797-887X
Adam L. Kraus  https://orcid.org/0000-0001-9811-568X

**References**

Afram, N., & Berdyugina1, S. V. 2015, A&A, 576, A34

APASS, AAVSO Photometric All-Sky Survey 2020, https://www.aavso.org/apass

Astropy Collaboration 2018, AJ, 156, 123

Bailer-Jones, C. A. L, Rybizki, J., Fouesneau, M., Mantelet, G., & Andrae, R. 2018, AJ, 156, 58

Balona, L. A. 2015, MNRAS, 447, 2714

Bayo, A., Rodrigo, C., Barrado y Navascués, D., et al. 2008, A&A, 492, 277B

Belloni, T., Verbunt, F., & Mathieu, R. D. 1998, A&A, 339, 431

Bradley L, Sipocz B., Robitaille, T., Tollerud, E., et al. 2019, Zenodo, https://zenodo.org/record/2533376

CSS 2020, Catalina Sky Survey (Tucson, AZ: LPL), http://nesssi.cacr.caltech.edu/DataRelease/

Craine, E. R**.,** Tucker, R. A., Flurchick, K. M., Kraus, A. L., & Griego, B. F. 2013, SAS 32, Symposium on
    Telescope Science, ed. B. D. Warner, R. K. Buchheim, J. L. Foote, & D. Mais (Rancho Cucamonga, CA:
    SAS), 47

Da Costa, G. S., Freeman, K. C., Kalnajs, A. J., Rodgers, A. W., & Stapinski, T.E. 1977, AJ, 82, 810

Danielski, C., Babusiaux, C., Ruiz-Dern, L., Sartoretti, P., & Arenou, F. 2018, A&A, 614, A19

Davenport, J. R. A. 2015, IAU Symp. 320, Solar and Stellar Flares and their Effects on Planets, ed. A. G.
    Kosovichev, S. L Hawley, & P. Heinzel (Cambridge, UK: CUP), 128

Gaia Collaboration, Babusiaux, C., van Leeuwen, F., et al. 2018, A&A, 616, A10

Gaia Collaboration, Brown, A.G.A., Vallenari, A., et al. 2020, A&A, 649, A1

Gaia Collaboration, Helmi, A., van Leeuwen, F., et al. 2018, A&A, 616, A12

Gaia Collaboration, Prusti, T., de Bruijne, J. H. J., et al. 2016, A&A, 595, A1

Gansicke, B. T. 2016, ApJ, 818, L7

Geller, A. M, Leiner, E. M., Bellini, A., et al. 2017, ApJ, 840, 66

Gray R. O., Corbally, C. J., Garrison R. F., McFadden M. T., & Robinson, P. E. 2003, AJ, 126, 2048

Gray, R. O., & Corbally, C. J. 2009, Stellar Spectral Classification (Princeton, NJ: Princeton University Press)

Green, G. M., Schlafly, E. F., Finkbeiner, D., et al. 2018, MNRAS, 478, 651

Gunther, M. N., Zhuchang, Z., Seager, S., et al. 2020, AJ, 159, 60

Hawkins, K, Jofre, P., Masseron, T., & Gilmore, G. 2015, MNRAS, 453, 758

Hidalgo, S., Pietrinferni, A., Cassisi, S. et al. 2018, ApJ, 856, 125

Jayasinghe, T., Stanek, K. Z., Kochanek, C. S., et al. 2020, MNRAS, 493, 4186

Johnson, D. R. H., & Soderblom, D. R. 1987, AJ, 93, 864

Juric M., Ivezic, Z., Brooks, A., et al. 2008, ApJ, 673, 864

Karoff, C., Knudsen, M. F., De Cat, P., et al. 2016, Nat Commun, 7, 11058

Kovari, Zs., Olah, K., & Gunther, M. N. 2020, A&A, 641, A83

Kraus, A. L., Craine, E. R., Giampapa, M. S., Scharlach, W. W. G., & Tucker, R. A. 2007, AJ, 134, 1488

Kraus, A. L., Tucker, R. A., Thompson, M. I, Craine, E. R., & Hillenbrand, L.A. 2011, AJ, 728, 48

Lindegren, L., Hernández, J., Bombrun, A., et al. 2018, A&A, 616, A2

Luo, A.-L, Zhao, Y.-H, Zhao, G., et al. 2018, LAMOST Data Release 4 (Beijing, China: CAS), http://dr4.lamost.org/

Malamud, U., & Perets, H.B. 2020, MNRAS, 492, 5561

Mathieu, R. D., van den Berg, M., Torres, G., et al. 2003, AJ, 125, 246

Morris, B. M., Curtis, J. L., Sakari, C., Hawley, S. L., & Agol, E. 2019, AJ, 158, 101

Ramírez, I., & Meléndez, J. 2005, ApJ, 626, 465

Sandage, A., & Fouts. G. 1987, AJ, 93, 592

Shappee, B. J., Prieto, J. L., Grupe, D., et al. 2014, ApJ, 788, 48

Tonry, J., & Davis, M. 1979, AJ, 84, 1511

Tucker, R. A. 2007, AJ, 134, 1483

Tucker, R. A., Craine, E. R., Flurchick, K. M., Kraus, A. L., & Griego, B. F. 2013, SAS 32, Symposium on Telescope Science, ed. B. D. Warner, R. K. Buchheim, J. L. Foote, & D. Mais (Rancho Cucamonga, CA: SAS), 41

Wang, S. and Chen, X. 2019, ApJ, 877, 116

Zachrias, M. I., Finch, C. T., Girard, T. M., et al. 2013, AJ, 145, 44